\documentclass[12pt]{iopart} 

\newcommand\nice[1]{#1}    \newcommand\subm[1]{}   

\nice{\newcommand\dosingle[1]{#1}  \newcommand\dodouble[1]{ }} 
\subm{\newcommand\dosingle[1]{ }   \newcommand\dodouble[1]{#1}} 

\dodouble{\documentclass[usenatbib,referee]{mn2e}}

\usepackage{graphicx}

\usepackage{amsfonts}

\usepackage{natbib}  
\usepackage{url}

\newcommand\prerefereechanges[1]{#1}    

\newcommand\postrefereechanges[1]{#1}

 \usepackage{hyperref}

\nice{ 
\providecommand{\eprint}[1]{\href{http://arxiv.org/abs/#1}{{\tt [arXiv:#1]}}}

\providecommand{\url}[1]{\href{#1}{#1}}

}

\providecommand{\adsurl}[1]{} 
\newcommand\SSS{Sect.~}

\providecommand\apj{ApJ}                 
\providecommand\aap{A\&A}            
\providecommand\mnras{MNRAS}
\providecommand\cqg{CQG}

\providecommand\prd{Phys.~Rev.~D}
\providecommand\physrep{Phys. Rep.}

\providecommand\PRL{PRL}

\nice{ \newcommand\mycaptionfont{\protect\footnotesize} }

\newcommand\gtapprox{\,\lower.6ex\hbox{$\buildrel >\over \sim$} \, }
\newcommand\ltapprox{\,\lower.6ex\hbox{$\buildrel <\over \sim$} \, }
\newcommand\propapprox{\,\lower.6ex\hbox{$\buildrel \propto\over \sim$} \, }

\newcommand\arcs{\ifmmode {'' }\else $'' $\fi}     
\newcommand\arcm{\ifmmode {' }\else $' $\fi}       
\newcommand\ddeg{\ifmmode^\circ\else$^\circ$\fi}    

\newcommand\frtoday{Le\space\number\day\space\ifcase\month\or
  janvier\or f\'evrier\or mars\or avril\or mai\or juin\or
  juillet\or ao\^ut\or septembre\or octobre\or novembre\or 
d\'ecembre\fi\space \number\year}
\newcommand\todayISO{\number\year-\ifnum\month<10 0\fi\number\month-\ifnum\day<10 0\fi\number\day}

\providecommand\cqg{ClassQuantGra}   %
\newcommand\hGpc{\mbox{$h^{-1}$ Gpc}}
\newcommand\hMpc{\mbox{$h^{-1}$ Mpc}}

\newcommand\Omtot{\Omega_{\mathrm{tot}}} 

\newcommand\md{\mathrm{d}} 
\newcommand\me{\mathrm{e}} 

\providecommand\notea{^\mathrm{a}}
\providecommand\noteb{^\mathrm{b}}

\begin{document}
\title[Relativistic topological acceleration]{A relativistic model of the topological acceleration effect}

\author[Ostrowski, Roukema \& Buli\'nski]{Jan J. Ostrowski,
Boudewijn F. Roukema 
and Zbigniew P. Buli\'nski 
\\
Toru\'n Centre for Astronomy, Nicolaus Copernicus University,
ul. Gagarina 11, 87-100 Toru\'n, Poland
}


\date{\frtoday}




\newcommand\Nchainsmain{16}
\newcommand\Npergroup{four}

\begin{abstract}
{It has previously been shown heuristically that the topology of the
  Universe affects gravity, in the sense that a test particle near a
  massive object in a multiply connected universe is subject to a
  topologically induced acceleration that opposes the local attraction
  to the massive {object.}}
{{It} is necessary to check if this effect
  occurs in a fully relativistic solution of the Einstein equations
  that has a multiply connected spatial section.}
{A Schwarzschild-like exact solution that is multiply connected in one spatial
  direction is checked for analytical and numerical consistency with
  the heuristic result.}
{The T$^1$ (slab space) heuristic result is found to be
  relativistically correct. For a fundamental domain size of $L$, a
  slow-moving, negligible-mass test particle lying at distance $x$
  along the axis from the object of mass $M$ to its nearest multiple
  image, where $GM/c^2 \ll x \ll L/2$, has a residual acceleration
  away from the massive object of $4\zeta(3) G(M/L^3)\,x$, where
  $\zeta(3)$ is Ap\'ery's constant.  For $M \sim 10^{14} 
  M_\odot$ and $L \sim 10$ to $20{\hGpc}$, this linear expression is
  accurate to $\pm10\%$ over $3{\hMpc} \ltapprox x \ltapprox
  2{\hGpc}$.}
{Thus, at least in a simple example of a multiply connected universe,
  the topological acceleration effect is not
  an artefact of Newtonian-like reasoning, and its linear derivation
  is accurate over about three orders of magnitude in $x$.}
\end{abstract}


{}

\maketitle 



\newcommand\tlinearOK{
  \begin{table}
    \caption{\mycaptionfont Minimum and maximum distance $x$ for which
      the $4\zeta(3)(M/L^3)\,x$ approximation is accurate to within 10\%
      (cf. Fig.~\protect\ref{f-KNaccelrhozero}).${}{\notea}$
      \label{t-linearOK}}
    $$\begin{array}{l rrrrr} \hline\hline 
      \rule{0ex}{2.5ex} 
      M & 10^{-10} & 3\times 10^{-10} &  10^{-9} & 3\times 10^{-9} &  10^{-8} \\
      \hline
      \rule{0ex}{2.5ex}
      \log_{10}x_{\min} & -3.94  &  -3.77  &  -3.54  &  -3.37  &  -3.14  \\ 
      \log_{10}x_{\max} & -0.63  &  -0.63  &  -0.63  &  -0.63  &  -0.63  \\ 
      \Delta \log_{10}x{\noteb}  & 3.32  &  3.14  &  2.91  &  2.74  &  2.51  \\ 
      \hline
    \end{array} $$
    \\
    ${}{\notea}${For convenience, $L=1$.} \\
    ${}{\noteb}${Range in $\log_{10} x$.}
  \end{table}
}  

\newcommand\fslab{
\begin{figure}
\centering 
\includegraphics[width=80mm]{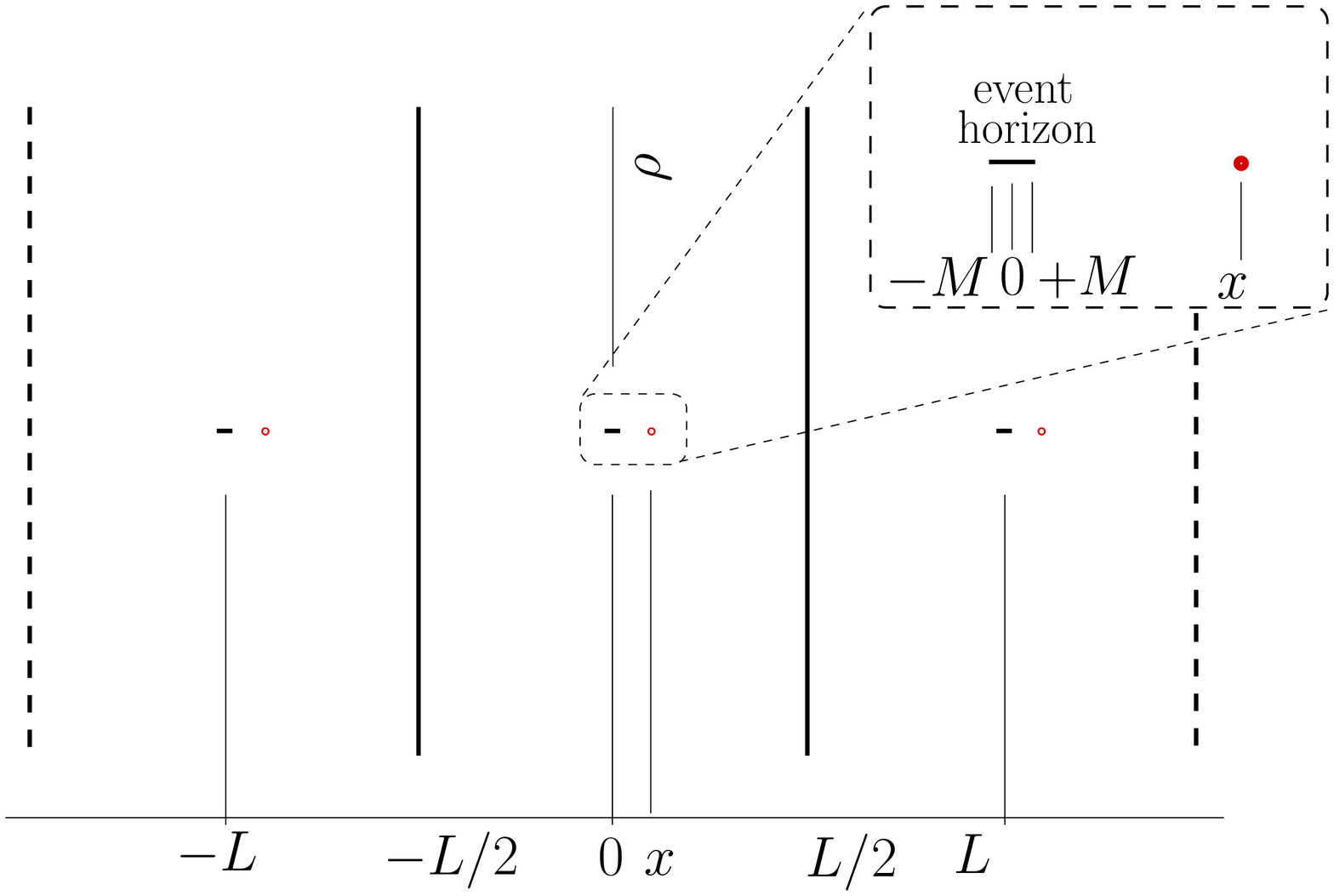} 
\caption[]{ \mycaptionfont { 
Event horizon of massive object of mass $M$ (dark line segment) and test particle 
(dot) in slab space T$^1$ of fundamental domain
length $L$. In a 3-section of constant coordinate time $t$, 
the event horizon of the massive object is
spherically symmetric in terms of the metric, 
but in the Weyl coordinates used here, it
is a line segment extending from $-M$ to $+M$ in the $x$ direction,
where $G=c=1$, and is of zero size in the radial coordinate $\rho$.
The test particle is displaced a small distance from the massive object,
i.e. $x \ll L/2$, but well outside the event horizon, i.e.
$0 < M \ll x$ (zoom) [Eq.(\protect\ref{e-x-assump})].
}
\label{f-slab}}
\end{figure} 
} 

\newcommand\fftwoaccelrhozero{
\begin{figure}
\centering 
\includegraphics[width=70mm]{F2_accel_rho0} 
\caption[]{ \mycaptionfont {
    {{Topological}}
    acceleration $\mathrm{d}^2 z /\mathrm{d}\tau^2$ as a function of
    distance $z$ from the massive object, along the spatial geodesic
    of length $2\pi$ joining the massive object of mass $\mu =
    10^{-7}$ to itself. The thick curve shows $\mathrm{d}^2 z
    /\mathrm{d}\tau^2$ using Eqs~(3), (63) and (66) as discussed in
    \SSS\protect\ref{s-result-Frolov}, the thin ascending line shows
    the first-order {{topological acceleration}} effect
    from Eq.~(7) of \protect\citet{RBBSJ06}, and the thin descending
    line shows the difference between the Newtonian and (simply
    connected) Schwarzschild accelerations (i.e. at $z \gtapprox
    z_{\mathrm{S}}$, where $z_{\mathrm{S}}$ is the Schwarzschild
    radius). {{\bf TODO: tell the reader where there's a discussion of
        the units.}}  }
\label{f-ftwoaccelrhozero}}
\end{figure} 
} 

\newcommand\fKNaccelrhozero{
  \begin{figure}
    \centering 
    \includegraphics[width=70mm]{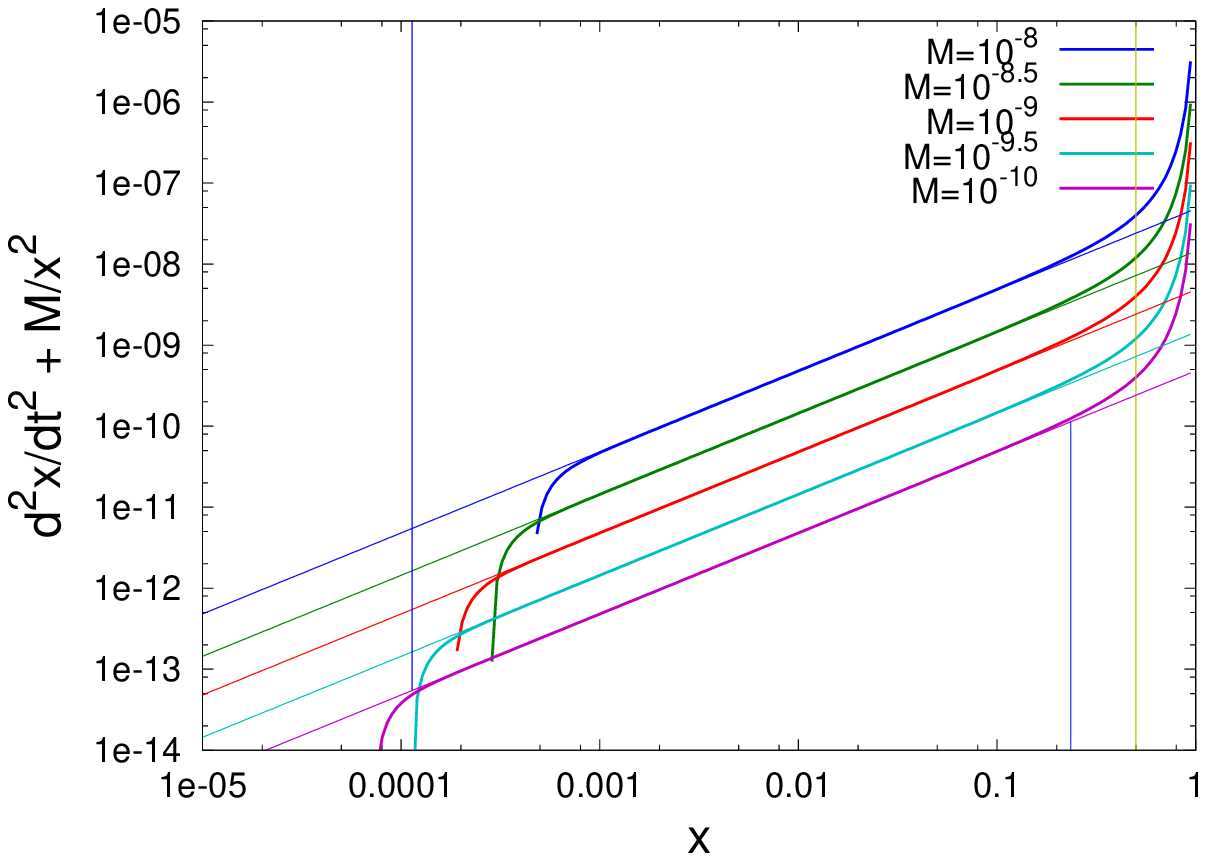} 
    \caption[]{ \mycaptionfont {{{Topological}}
        acceleration $\mathrm{d}^2 x /\mathrm{d}t^2 + M/x^2$ as a
        function of coordinate distance $x$ from the ``first'' copy of
        the massive object of mass $M$, along the spatial geodesic
        joining the massive object to itself, using
        Eqs~(\protect\ref{e-defn-vareps_0}),
        (\protect\ref{e-defn-omega_0}), and
        (\protect\ref{e-defn-vareps}) in
        Eq.~(\protect\ref{e-coordcoord-accel-step2}), for $L=1$. The
        curves show masses from $M=10^{-10}$ to $10^{-8}$, from bottom
        to top, respectively, as labelled. The sloping lines show
        $4\zeta(3) (M/L^3) x$. The vertical line at $x=0.5$ shows the
        halfway point to the ``next'' copy of the massive object.  The
        vertical lines down to and up to the $M=10^{-10}$ curve show
        the range over which the linear approximation is valid to
        within $\pm 10\%$.
        \label{f-KNaccelrhozero}
    }}
  \end{figure} 
} 


\section{Introduction}  \label{s-intro}

The topology of spatial sections of the Universe has been of interest
since the foundations of relativistic cosmology 
\citep[e.g., ][]{deSitt17,Fried23,Fried24,Lemaitre31ell,Rob35}\footnote{See \citet{Lemaitre31elltrans} for an
incomplete translation of \citet{Lemaitre31ell}.}.
{However, astronomical measurements aimed at
  determining spatial topology have long suffered a theoretical
  disadvantage compared to curvature measurements. The latter are
  tightly related to another type of astronomical measurement
  (matter-energy density) via a well-established physical theory: 
  general relativity. Some elementary work that might lead to 
  a physical theory of cosmic topology has been carried out, by
  comparing certain characteristics of different manifolds
  \citep{Masafumi96,CarlipSurya04} and by explorations of some
  elements of topology change in quantum gravity \citep[e.g.,
  ][]{DowS98}, but these remain very distant from astronomical
  observations. Thus, most of the empirical work of the past few decades
  has been carried out with no theoretical constraints, apart
  from assuming Friedmann-Lema\^{\i}tre-Robertson-Walker (FLRW) cosmological
  models.}
For recent
empirical analyses of the case for and against a multiply connected
positively curved spatial section, in particular the Poincar\'e dodecahedral space
S$^3/I^*$, see \citet{RK11} and references therein. 
{Theoretical developments could offer the possibility
of new astronomical tests for cosmic topology.}

For perfectly homogeneous solutions of the Einstein equations, i.e.
{FLRW} models, if the sign of the curvature is determined empirically,
only the covering space H$^3$, $\mathbb{R}^3$, or S$^3$, i.e. an
apparent space containing many copies of the fundamental domain, can
be inferred.  The 3-manifold of comoving space itself, i.e.
H$^3/\Gamma$, $\mathbb{R}^3/\Gamma$, or S$^3/\Gamma$, respectively,
for a fundamental group (holonomy group) $\Gamma$, {is not implied by
  the curvature}. However, heuristic, Newtonian-like calculations have
recently shown that the presence of a single inhomogeneity is enough
to, in principle, distinguish different possible spaces of the same
fixed curvature \citep{RBBSJ06,RR09}. The effect is a long-distance,
globally induced acceleration term that opposes the local attraction
to a massive object. Curiously, the effect, (hereafter, ``topological
acceleration''\protect\footnote{The implied meaning is ``acceleration
  induced globally by cosmic topology''; a term used in earlier work
  is ``residual gravity''.})  is much weaker in the Poincar\'e
dodecahedral space than in other spaces, hinting at a theoretical
reason for selecting this space independent of the observational
reasons published six years \prerefereechanges{earlier.}
\postrefereechanges{Other calculations of globally induced effects of
  the spatial topology of a fixed FLRW background include
  \citet{InfeldSchild46} and \citet{Bernui98raddamping}.}

Astronomical tests for cosmic topology are almost always based on the
fact that photons from a point in comoving space can arrive at the
observer by multiple paths \citep[e.g. ][]{2002nmgm.meet.1937R}.  In
contrast, topological acceleration has the potential to provide tests
of galaxy kinematics that could provide experimental constraints on
cosmic topology independent of tests based on multiple imaging (of
extragalactic objects or of cosmic microwave background temperature
fluctuations).  However, the calculations of the effect
\citep{RBBSJ06,RR09} did not use full solutions of the
Einstein equations. The FLRW metric is an exact solution of the 
Einstein equations, but it
can only by applied to the real Universe as a heuristic guide, since
galaxies (inhomogeneities) exist. When this heuristic guide is 
used to interpret recent observational constraints, an accelerating
scale factor and dark energy are inferred. That is, the latter appear to 
be artefacts of a heuristic approach \citep[e.g.][]{CBK10,BuchCarf03,WiegBuch10}.
Similarly, since a heuristic, Newtonian-like approach was
used to calculate the topological acceleration effect, the latter could,
in principle, also be an artefact. Thus, it is
important to check 
whether the effect
really exists in the relativistic case of a $(3+1)$-dimensional spacetime
with multiply connected spatial sections.
\postrefereechanges{Indeed, in reviews of cosmic topology
  \citep[e.g.,][]{Ellis1971,LaLu95,LR99,LevinReview2002,RG04}, the use
  of the exact (locally homogeneous) FLRW models has generally led to
  the inference that in a classical general relativistic model,
  spatially global topology should only constrain the 
  curvature-related parameters of the metric \citep[for methods, see
    e.g.,][]{ULL99a,Roukema99-1,Bento06PoincareOmegak,Reboucas06Omegak,Bento06Chaplygin,Reboucas06Omk}.
  This inference only becomes incorrect when the universe model
  contains at least one inhomogeneity.}

{Strictly speaking, from the point of view of
  differential geometry alone, astronomical tests for cosmic topology
  based on multiple imaging are degenerate.  They do not distinguish a
  multiply connected space from a simply connected universal cover
  that happens to be populated by physically distinct objects located
  at appropriate positions in a perfectly regular tiling.\footnote{{If
    a spatially closed, timelike path (world-line) were allowed around
    a non-contractible spatial loop $S^1$ in a given universe model,
    then a physical voyage by an observer around the loop would
    provide an experimental method to overcome this degeneracy.}}  Similarly, what we
  refer to here as ``topological acceleration'' could be
  interpreted as an effect of a perfectly regular tiling in a simply
  connected covering space.  In this paper, we adopt the
  Schwarzschildian approach to lifting the degeneracy, ``We would be
  much happier with the view that these repetitions are illusory, that
  in reality space has peculiar connection properties so that if we
  leave any one cube through a side, then we immediately reenter it
  through the opposite side,''  \protect\citep[transl.,][]{Schw00,Schw98} This is
  similar to the interpretation of what are normally claimed to be
  multiple, ``gravitationally lensed'' images of a single physical object 
  as representing a single object rather than
  multiple objects that happen to mimic what is expected from a 
  gravitational lensing hypothesis \citep[e.g.,][]{AdamBacon89}.}

Here, we compare the heuristic calculation of \citet{RBBSJ06} for
T$^1$ (slab space, i.e. S$^1 \times \mathbb{R}^2$) to a limiting case
of an exact, Schwarzschild-like solution which has a T$^1$ spatial
section outside of the event horizon.  {This is a
  simpler 3-manifold than those that seem to be favoured by
  observations \citep[as stated above, see][and references
    therein]{RK11}, with only one short dimension, and it only
  contains one massive object. Checking the existence of topological
  acceleration in this relativistic case does not guarantee
  corresponding results in more complicated cases, but it does open
  the prospect of generalisations.}  In \SSS\ref{s-method-assump}
we list the simplifying assumptions for considering a test particle
distant from the event horizon but much closer to the ``first'' copy
of a massive object than to an ``adjacent'' copy. The metric in Weyl
coordinates and related functions \citep{KN94} are given in
\SSS\ref{s-method-KN} for an analytical derivation and the numerical
method is stated in \SSS\ref{s-result-numer} The analytical and
numerical calculations are presented in \SSS\ref{s-result-KN} and
\SSS\ref{s-result-numer}, respectively,
and a conclusion is
given in \SSS\ref{s-conclu}.

\fslab

\section{Method} \label{s-method}
For the Newtonian-like calculation, 
Fig.~3 of \citet{RBBSJ06} illustrates the situation of a test particle
at distance $x$ from
a point object of mass $M$ in T$^1$, lying along the spatial geodesic of
length $L$ from the
massive object to itself. 
Defining $\epsilon := x/L$ gives
the topologically induced acceleration \citep[Eq.~(7),][]{RBBSJ06}
\begin{eqnarray}
\ddot{x} &=& M \sum_{j=1}^{\infty} 
  \left[ \frac{1}{(jL-x)^2} - \frac{1}{(jL+x)^2}  \right]
\approx
  4.8  \frac{M \epsilon}{L^2}  
\label{e-newton-like}
\end{eqnarray}
when $\epsilon \ll 1$, adopting $G=1$. 

\subsection{Assumptions} \label{s-method-assump}
A Schwarzschild-like solution of the Einstein equations has an event
horizon, but behaviour close to the event horizon (or inside it) is
not of interest here. Also, if the {topological
  acceleration} effect were to be analysed for its effects on galaxy
dynamics, peculiar velocities would typically be of the order of
$10^{-3}$c and very rarely exceed $10^{-2}$c. As in the earlier
derivation of Eq.~(\ref{e-newton-like}), we consider a test particle
lying along the closed spatial geodesic joining the massive object to
itself, i.e. having Weyl radial coordinate $\rho=0$.

Thus, in addition to adopting the conventions $G=c=1$,
a $(-,+,+,+)$ metric, Greek spacetime indices $0,1,2,3$ 
and Roman space indices $1,2,3$, 
our assumptions become:
\begin{equation} 
  0 < M \ll x \ll L/2 ,
  \label{e-x-assump}
\end{equation}
\begin{equation} 
  \frac{\mathrm{d}x^i }{ \mathrm{d}t} \ll 1 
\;\;
\Rightarrow \;\;
  \frac{\mathrm{d}x^i }{ \mathrm{d}\tau} \ll 
  1 \ltapprox
  \frac{\mathrm{d}t}{\mathrm{d}\tau } ,
  \label{e-low-beta}
\end{equation}
\begin{equation} 
  \rho = 0 ,
  \label{e-zero-rho}
\end{equation}
where $\tau$ is the proper time along the test particle's world line
$x^\alpha(\tau)$, and $t \equiv x^0$. 
{The assumption of a low coordinate velocity
implies a low 4-velocity spatial component
[Eq.~(\ref{e-low-beta})].}

\subsection{Analytical approach} \label{s-method-KN}
\citet{KN94} presented a family of multiply connected exact
solutions of the Einstein equations that includes
a Schwarzschild-like solution with T$^1$ spatial sections outside
of the event horizon.
Figure~\ref{f-slab} shows this solution in Weyl coordinates $x$ and $\rho$.
Using Weyl coordinates and the Ernst potential, the
metric is
\begin{equation}
  \md s^2 = -\me^{\omega} \md t^2 +
  \me^{-\omega} \left[ \me^{2k} (\md x^2 + \md \rho^2) + \rho^2 \md \phi^2 \right]
  \label{e-KN-metric}
\end{equation}
from Eq.~(9) of \citet{KN94},\footnote{In the online version 1 of {\tt
    arXiv:gr-qc/9403029}, the time component has an obvious
  typological error in the sign.}  where $k$ and $\omega$ relate to an
Ernst potential $\varepsilon$ defined on $\xi := x + \mathrm{i} \rho$,
with
\begin{equation}
  \varepsilon_0(x,\rho) := 
  \frac{ \sqrt{(x-M)^2 +\rho^2} + \sqrt{(x+M)^2 +\rho^2} -2M}{
   \sqrt{(x-M)^2 +\rho^2} + \sqrt{(x+M)^2 +\rho^2} +2M}
  \label{e-defn-vareps_0} 
\end{equation}
\begin{equation}
  \omega_0 := \ln \varepsilon_0,\; a_0 := 0,\; 
  {{{a_{j\not=0} := \frac{2M}{L\,|j|}}}}
  \label{e-defn-omega_0}
\end{equation}
\begin{equation}
  \omega(x,\rho) := \sum_{j=-\infty}^{\infty} \left[\omega_0(x+jL,\rho) + a_j\right]
  \label{e-defn-omega}
  ,\;\;\;
  \varepsilon := \me^\omega
  \label{e-defn-vareps}
\end{equation}
\begin{equation}
  \partial_\xi k = 2 \mathrm{i} \rho 
  \frac{\partial_\xi \varepsilon \partial_\xi \bar{\varepsilon}}{
    (\varepsilon + \bar{\varepsilon})^2},
  \label{e-defn-k}
\end{equation}
from Eqs~(13), (12), (5) and (7) of \cite{KN94}.\footnote{Positive square roots
are implied in Eq.~(\ref{e-defn-vareps_0}).}
 The Ernst potential is
real for this solution, so Eq.~(\ref{e-defn-k}) gives
\begin{equation}
  \partial_\xi k 
  = 
  \frac{1}{2} \mathrm{i} \rho 
  \frac{(\partial_\xi \varepsilon)^2}{\varepsilon^2} 
  = 
  \frac{1}{2} \mathrm{i} \rho 
  \left[ \partial_\xi (\ln \varepsilon) \right]^2 
  = 
  \frac{1}{2} \mathrm{i} \rho (\partial_\xi \omega)^2.
  \label{e-k-CSchw}
\end{equation}
Thus, at $\rho=0$, we have $\partial_\xi k =0$, i.e. $k$ is a 
constant along $\rho=0$. We choose $k(\rho=0)=0$, since any other
value just implies a rescaling of units. Thus, the metric on $\rho=0$ is
\begin{equation}
  \md s^2 = -\me^{\omega} \md t^2 +
  \me^{-\omega} \left( \md x^2 + \md \rho^2 + \rho^2 \md \phi^2 \right) .
  \label{e-KN-metric-rho-0}
\end{equation}

Defining the tangent vector $\vec{V}$ by components $V^\alpha := \md
x^\alpha/\md \tau$, the geodesic equation for the test particle is
$\nabla_V \vec{V} = \vec{0}$, i.e.
\begin{equation}
  \frac{\md^2 x^\alpha}{\md \tau^2} + 
  \Gamma^\alpha_{\;\;\beta\gamma} 
  \frac{\md x^\beta}{\md \tau} 
  \frac{\md x^\gamma}{\md \tau} = 0,
  \label{e-geodesic-general}
\end{equation}
where $\Gamma^\alpha_{\;\;\beta\gamma}$ are the Christoffel symbols of the second
kind.\footnote{Symmetric definition.} 
{Since a $(-,+,+,+)$ convention is used  and low velocities are assumed
[Eq.~(\ref{e-low-beta})],} 
coordinate and proper
time are related by 
\begin{equation}
  \md \tau^2 = |g_{00}| \, (\md x^0)^2  = e^\omega \, \md t^2 ,
  \label{e-t-tau}
\end{equation}
giving the $x^1$ coordinate acceleration
\begin{equation}
  \frac{\md^2 x^1}{\md t^2} = 
  \frac{\md^2 x}{\md t^2} = 
e^{-\omega} 
  \frac{\md^2 x}{\md \tau^2} .
  \label{e-coordcoord-accel-step1}
\end{equation}
Given the conditions in Eqs.~(\ref{e-x-assump}), 
(\ref{e-low-beta}), and (\ref{e-zero-rho}) and the Ernst potential expressions in
Eqs.~(\ref{e-KN-metric-rho-0}), (\ref{e-defn-vareps_0}),
  (\ref{e-defn-omega_0}), and
  (\ref{e-defn-vareps}), conversion to metric units of $x$ and $t$ would only
modify this to second order. 
In \SSS\ref{s-result-KN}, {${\md^2 x}/{\md t^2}$ is evaluated.}

\subsection{Numerical approach} \label{s-method-numer}
The Ernst potential expressions [Eqs~(\ref{e-defn-vareps_0}),
  (\ref{e-defn-omega_0}), and (\ref{e-defn-vareps})] are used for
numerical evaluation of Eq.~(\ref{e-coordcoord-accel-step2}) (below)
by finite differencing over small intervals.
A typical cluster of galaxies 
has a mass of $\sim 10^{14} M_\odot$, i.e. $M = 4.78$~pc in length
units, and the most massive clusters have masses up to about $10^{15}
M_\odot$.\footnote{In the Weyl coordinate $x$, the event horizon
  occurs along $|x| \le M,$ $\rho = 0$ (Fig.~\protect\ref{f-slab}, which
  may seem surprising given that the event horizon in Schwarzschild
  coordinates occurs at $r=2M$.  See
  \protect\citet[e.g.,][Sect.~II.B. and references therein]{Frolov03}
  for the relation between Weyl and Schwarzschild coordinates.}
Observational estimates of $L$ are in the range 10 to 20{\hGpc}
\citep[e.g.][and references therein]{RK11}. Thus, typical scales of
$M/L$ of interest should be $M/L \sim 10^{-10}$ to $10^{-8}$.
Since $M/x, M/L,$ and $x/L$ are small, the 
{use of 53-bit--significand}
double-precision floating-point numbers is replaced
by arbitrary precision arithmetic, for $10^{-10} < M/L < 10^{-8}$.
These calculations are presented in \SSS\ref{s-result-numer}.

\section{Results} \label{s-results}

\subsection{Coordinate acceleration}
\label{s-result-KN}

Given that 
the test particle is far from the ``close'' copy of the 
event horizon and from ``distant'' copies
[Eq.~(\ref{e-x-assump})], we have $| \Gamma^\alpha_{\;\;\beta\gamma} | \ll 1$.
Together with Eq.~(\ref{e-low-beta}), this implies that most
of the terms in the sum in Eq.~(\ref{e-geodesic-general}) are small terms
of second order or higher, leaving
\begin{equation}
  \frac{\md^2 x^\alpha}{\md \tau^2} + 
  \Gamma^\alpha_{\;\;00} 
  \left(\frac{\md x^0}{\md \tau} \right)^2
 \approx 0.
  \label{e-geodesic-0}
\end{equation}
Since the metric is static (and diagonal), we have
\begin{equation}
  \Gamma^\alpha_{\;\;00} = \frac{1}{2} g^{\alpha\lambda} 
  (\partial_0 g_{\lambda 0} + \partial_0 g_{0\lambda} - \partial_\lambda g_{00})
  = -\frac{1}{2} g^{\alpha\lambda}\partial_\lambda g_{00}.
  \label{e-Gamma-simplify}
\end{equation}
Using Eq.~(\ref{e-t-tau}), 
Eq.~(\ref{e-geodesic-0}) becomes
\begin{equation}
  \frac{\md^2 x^\alpha}{\md \tau^2} =
  \frac{1}{2} 
  \frac{g^{\alpha\lambda}}{|g_{00}|}  \partial_\lambda g_{00}.
  \label{e-coord-accel-general}
\end{equation}
Since the metric
is diagonal, we have $g^{xx} = (g_{xx})^{-1} = e^{+\omega}$ and 
\begin{equation}
  \frac{\md^2 x^1}{\md \tau^2} =
  \frac{1}{2}   \frac{g^{11}}{|g_{00}|}  \partial_1 g_{00},
\end{equation}
i.e.,
\begin{equation}
  \frac{\md^2 x}{\md \tau^2} =  
  \frac{1}{2}   \frac{g^{xx}}{|g_{tt}|}  \partial_x g_{tt} 
  =
  \frac{1}{2}   \frac{e^\omega}{e^\omega}  (-\partial_x e^\omega) 
  =
  - \frac{1}{2} e^\omega   \partial_x \omega.
  \label{e-coord-accel}
\end{equation}
Thus, from Eq.~(\ref{e-coordcoord-accel-step1}),
\begin{equation}
  \frac{\md^2 x}{\md t^2} =  -\frac{1}{2} \partial_x \omega.
  \label{e-coordcoord-accel-step2}
\end{equation}
{
Substituting Eq~(\ref{e-defn-omega_0}) 
in Eq.~(\ref{e-defn-vareps}a) 
[corresponding to (12) and (5) of \cite{KN94}, respectively]
\begin{eqnarray}
  \omega(x,\rho) 
  &=& w_0(x,\rho) + 
  \sum_{j=-\infty,j\not=0}^{\infty} 
  \left[ \omega_0(x+jL,\rho) + \frac{2M}{L\,|j|} \right].
  \label{e-omega}
\end{eqnarray}
This is the convergent solution found by \cite{KN94}.}
Dropping the $\rho$ dependence since we are interested in the $\rho=0$ axis,
the derivative can be written using Eq.~(\ref{e-defn-omega_0})
\begin{eqnarray}
  \partial_x \omega(x) 
  &=& 
  \partial_x 
  \left[ \ln \varepsilon_0(x) + \sum_{j=-\infty,j\not=0}^{\infty} 
    \ln \varepsilon_0(x+jL)
    \right]
  \nonumber \\
  &=& 
  \frac{\partial_x \varepsilon_0(x)}{\varepsilon_0(x)} + 
  \sum_{j=1}^{\infty} \frac{\partial_x \varepsilon_0(x+jL)}{\varepsilon_0(x+jL)} + 
  \sum_{j=1}^{\infty} \frac{\partial_x \varepsilon_0(x-jL)}{\varepsilon_0(x-jL)} 
  \nonumber \\
  &=& 
  \frac{\partial_x \varepsilon_0(x)}{\varepsilon_0(x)} + 
  \sum_{j=1}^{\infty} \frac{\partial_x \varepsilon_0(x+jL)}{\varepsilon_0(x+jL)} - 
  \sum_{j=1}^{\infty} \frac{\partial_x \varepsilon_0(jL-x)}{\varepsilon_0(jL-x)} 
  \nonumber \\
  &\approx&
  {\partial_x \varepsilon_0(x)} +
  \sum_{j=1}^{\infty} {\partial_x \varepsilon_0(jL+x)} - 
  \sum_{j=1}^{\infty} {\partial_x \varepsilon_0(jL-x)},
  \nonumber 
  \\
\end{eqnarray}
where the third equality follows from using
$\partial_{x'}\varepsilon_0(-x') = - \partial_{x'}
\varepsilon_0(x')$ [cf. Eq.~(\protect\ref{e-defn-vareps_0})], and the
approximation follows from using Eqs~(\protect\ref{e-x-assump}),
(\protect\ref{e-zero-rho}), and $j\ge 1$ in
Eq.~(\protect\ref{e-defn-vareps_0}).

\fKNaccelrhozero

\tlinearOK

For $0 < M < x'$, Eq.~(\ref{e-defn-vareps_0}) simplifies to 
\begin{equation}
  \varepsilon_0(x',0) = \frac{x'-M}{x'+M}.
  \label{e-vareps_0-simp}
\end{equation}
Again defining $\epsilon := x/L$, we then have 
\begin{eqnarray}
  &&\rule{-3ex}{0ex}\partial_x \omega(x) \nonumber \\
  &\approx&
  \frac{2M}{(x+M)^2} + 
  \sum_{j=1}^{\infty} \frac{2M}{(jL+x+M)^2} -
  \sum_{j=1}^{\infty} \frac{2M}{(jL-x+M)^2}
  \nonumber \\
  &\approx& 
  \frac{2M}{x^2} + 
  \frac{2M}{L^2} 
  \left(
  \sum_{j=1}^{\infty} \frac{1}{(j+\epsilon)^2} -
  \sum_{j=1}^{\infty} \frac{1}{(j-\epsilon)^2}  
  \right)
  \nonumber \\
  &\approx&
  \frac{2M}{x^2} + 
  \frac{2M}{L^2} 
  \left[
  \sum_{j=1}^{\infty} \frac{1}{j^2} \left( 1 - \frac{2\epsilon}{j} \right) -
  \sum_{j=1}^{\infty} \frac{1}{j^2} \left( 1 + \frac{2\epsilon}{j} \right) 
  \right]\rule{-3ex}{0ex}
  \nonumber \\
  &=& 
  \frac{2M}{x^2} - 
  \frac{8M}{L^2} \epsilon \sum_{j=1}^{\infty} \frac{1}{j^3} 
  \nonumber \\
  &=&
  \frac{2M}{x^2} - 
  \frac{8\zeta(3) M}{L^2} \epsilon
  \label{e-domegadx-step2}
\end{eqnarray}
where Eq.~(\ref{e-x-assump}) is applied in the second and third lines, and
$\zeta(3)$ is the Riemann zeta function evaluated at 
\postrefereechanges{an argument of 3}, 
i.e. Ap\'ery's constant.
  
Hence, using Eq.~(\ref{e-coordcoord-accel-step2}),
\begin{equation}
  \frac{\md^2 x}{\md t^2} \approx 
  - \frac{M}{x^2} + 
  \frac{4\zeta(3) M}{L^2} \, {\frac{x}{L}} 
  \approx 
  - \frac{M}{x^2} + 
  \frac{4.8 M}{L^2} \, {\frac{x}{L}} \; 
  \label{e-coordcoord-accel-step3}
\end{equation}
i.e. the coordinate acceleration is the Newtonian acceleration towards
the local copy of the massive object plus an opposing
{topological acceleration} term matching
Eq.~(\ref{e-newton-like}), i.e.  the expression derived earlier in
\citet{RBBSJ06}.

\subsection{Numerical check}
\label{s-result-numer}
For $10^{-10} < M < 10^{-8}$, $L=1$, {$|j| \le 14$,} 
and linear offsets of $\delta
x = \pm 10^{-13}$ around each $x$ value, numerically convergent
results were found {using 100 bits} in the
significand. Figure~\ref{f-KNaccelrhozero} and Table~\ref{t-linearOK}
show that in this case, the linear approximation for the residual
acceleration, $4\zeta(3) (M/L^3) \,x$, is accurate to within 10\% over
about three orders of magnitude.  Near the halfway point and at
$x>0.5$, i.e. closer to the ``next'' image of the massive object than
the ``local'' copy, the residual term defined in relation to the
acceleration towards the ``local'' copy is clearly no longer
small---it diverges as the test particle approaches the ``next''
copy.
Thus, for $L\sim 10$ to $20${\hGpc}, the linear expression should be accurate 
on scales of 
$3{\hMpc} \ltapprox x \ltapprox 2{\hGpc}$ in a slab-space universe containing
just one cluster of {about $10^{14} M_\odot$.}


\newcommand\dropthediscussion{
\section{Discussion} \label{s-disc}

{We have found that the topological acceleration effect is
  relativistically valid in (at least) the Schwarzschild-like slab
  space. Does this have implications regarding the
  accuracy of $N$-body simulations of the gravitational collapse of
  large-scale structure and the cosmic web? \citet{Melott90} briefly
  showed that density fluctuation evolution on the largest scales of
  an $N$-body simulation is affected by cosmic topology, at least in
  the two-dimensional case. Here we are concerned with the
  accelerations on individual particles, which can be used as a first
  step towards understanding the effect found by \citet{Melott90}.

  $N$-body simulations \citep[e.g., ][and references
    therein]{Bagla05,Dehnen11} are generally made assuming a T$^3$ model for
  comoving space, typically with a universe size $L$ that is smaller
  than is observationally realistic. The algorithms are typically
  developed with the aim of considering large-scale effects induced by
  the global topology to be artefacts, i.e. the aim is to ignore the
  physical effects of cosmic topology.  The most frequently used
  approach is to use Fourier techniques either explicitly, with a
  particle mesh, or implicitly, using Ewald summation in a tree code
  \citep[e.g. ][]{HernquistBS91}.  In \citet{RBBSJ06} and \citet{RR09}
  it was shown that for the Newtonian-like calculation of topological
  acceleration, the linear terms induced by topological images in
  different directions cancel each other exactly in a regular T$^3$
  model, leaving topological acceleration that is dominated by a term
  that is cubical in $x/L$. In a T$^3$ model of unequal side lengths,
  a linear term in the shortest compact direction dominates.

  Thus, to the degree to which they are numerically accurate, the most
  frequently used $N$-body methods eliminate the linear $x/L$
  topological acceleration term, leaving cubical $(x/L)^3$ terms.
  There has recently been considerable observational interest in T$^3$
  candidate FLRW models
  {\protect\citep{WMAPSpergel,Aurich07align,Aurich08a,Aurich08b,Aurich09a,AslanMan11}},
  so for this case, standard $N$-body simulations could be analysed in
  detail in order to study the expected effects of topological
  acceleration on galaxy kinematics, in the nearly-Newtonian
  approximation. Since the effect is anisotropic, alignments with the
  fundamental directions ($x,y,z$ axis directions of the simulation
  box) and vectorial calculations \citep{RBBSJ06} could be used to
  distinguish topological effects from numerical effects.

  On the other hand, except for well-proportioned FLRW models
  \citep{WeeksWellProp04}, the unequal lengths of comoving closed
  spatial geodesics in different directions imply that the linear term
  in $x/L$ should dominate in a ``typical'' multiply connected FLRW model. Standard
  $N$-body simulation methods would need to be modified in order to
  study topological acceleration effects in these cases.  Since the
  standard methods are based on either explicit or implicit Fourier
  techniques, relatively minor modifications of these might be enough
  to study both T$^1$ slab spaces and the T$^2$ model favoured by
  \citet{AslanMan11}.

  For FLRW models with even a small amount of curvature (e.g. a total
  density parameter of $\Omtot = 1.015$), especially the Poincar\'e
  dodecahedral space $S^3/I^*$
  \citep{LumNat03,Aurich2005a,Aurich2005b,Gundermann2005,Caillerie07,RBSG08,RBG08},
  standard $N$-body simulations are incorrect for several reasons.
  Firstly, $\Omtot \not=1$ does not only affect scale factor evolution
  $a(t)$, it also corresponds to curved comoving space. Secondly, 
  the fundamental domain is, in general, not cubical, and even in cases
  where a cubical fundamental domain can be used 
  \citep{Aurich11cube}, the identification of faces is different from
  that in the T$^3$ model normally assumed for $N$-body simulations.
  Thus, Fourier techniques cannot be applied, although they could be
  replaced by harmonic analysis that uses orthonormal sets of basis
  functions that are less trivial than the sinusoidal basis functions
  used in Fourier techniques.
  
  In contrast to models in which the acceleration is linear, i.e. 
  stronger than in T$^3$, both the
  linear and cubical terms cancel in the Poincar\'e
  space \citep{RR09}, leaving a dominant fifth order term.
  In this case, correctly designed $N$-body simulations 
  should show topological acceleration effects that are {\em weaker}
  than in the standard simulations.

  Thus, in order for $N$-body simulations to be used to develop
  methods of constraining the topology of the Universe independent of
  multiple imaging methods, the simulation design and algorithms would
  need to be modified in ways that should lead to topological
  acceleration that in ``typical'' FLRW models
  should be stronger than for the T$^3$ case, and in the Poincar\'e 
  dodecahedral space weaker than for T$^3$.}
}


\section{Conclusion} \label{s-conclu}

The T$^1$ (slab space) heuristic result is found to exist in the limit
of the Schwarzschild-like, exact, slab-space solution of the Einstein equations 
found by \citet{KN94}, with the same linear
{topological acceleration} term 
as for the simpler calculation.  For a fundamental domain 
{size of $L$,}
a slow-moving [Eq.~(\ref{e-low-beta})],
low-mass test particle lying at distance $x\ll L/2$ along the axis [$\rho =0$, Eq.~(\ref{e-zero-rho})]
from the object of positive mass $GM/c^2 \ll x$ [Eq.~(\ref{e-x-assump})] to its nearest
multiple image, the additional acceleration away from the
massive object is $4\zeta(3) G(M/L^3)\,x$
[Eq.~(\ref{e-coordcoord-accel-step3})].
For 
{a topological acceleration term}
away from a cluster of $\sim 10^{14} M_\odot$ 
and a Universe of size\footnote{More formally,
twice the injectivity radius, or twice the in-radius \protect\citep[Fig.~10,][ and references therein]{LR99}.}
$L\sim 10$ to $20${\hGpc}, the linear expression should be accurate 
over $3{\hMpc} \ltapprox x \ltapprox 2{\hGpc}$ 
(Fig.~\ref{f-KNaccelrhozero}, 
{column with $M = 3\times 10^{-10}$ in} Table~\ref{t-linearOK}).
Thus, at least in a simple example of a multiply connected universe,
the 
{topological acceleration}
effect is not an artefact of Newtonian-like
reasoning. The linear term exists as an analytical limit of the relativistic model 
and as a good numerical approximation to it
{over several orders of magnitude of $x$.}

This qualitatively suggests that the Newtonian-like derivations of 
the 
{topological acceleration}
effect for well-proportioned FLRW models 
\citep{RR09}, in which the Poincar\'e dodecahedral space is found
to be uniquely well-balanced, might also be valid limits of fully
relativistic Schwarzschild-like solutions with these spatial 
sections. Other extensions of the present work within 
the slab-space solution could include consideration 
of test particles with high coordinate velocities 
{[violating Eq.~(\ref{e-low-beta})],} or expanding (FLRW-like) solutions.

{Although we have shown that topological
  acceleration exists theoretically, the effect is clearly weak over
  the range for which the linear expression is accurate.  It is not
  clear how easy it might be to test the effect observationally.  A
  possible method might follow from considering the effect as a
  feedback effect from global anisotropy to local anisotropy. Multiply
  connected FLRW models have usually been thought to be {\em locally}
  isotropic (the metric on a spatial section is isotropic) and {\em
    globally} anisotropic (some global tests differ depending on the
  chosen spatial direction). Here, we have confirmed that global
  anisotropy can imply a local acceleration effect. The effect is
  clearly anisotropic. The integration of this weak effect over a
  long period of time for individual particle trajectories or 
  collapsing density perturbations might, in
  principle, give anisotropic long-term statistical behaviour that is
  strong enough to be detectable.
  Thus, the global
  anisotropy of a multiply connected space implies (at least in the
  case considered here) a weak, locally anisotropic dynamical
  effect. Whether or not the time-integrated effect could be strong enough to detect
  in observations of galaxy kinematics or cosmic web structure is 
  a question worth exploring in future work.}


\ack
\sloppy 
Thank you to several people for useful anonymous comments.
Use was made of the Centre de Donn\'ees astronomiques de Strasbourg 
(\url{http://cdsads.u-strasbg.fr}), 
the GNU multi-precision (GMP) and MPFR libraries,
and
the GNU {\sc Octave} command-line, high-level numerical computation software 
(\url{http://www.gnu.org/software/octave}). 
\postrefereechanges{A part of this project has made use of 
Program Oblicze\'n Wielkich Wyzwa\'n nauki i techniki (POWIEW)
computational resources (grant 87) at the Pozna\'n 
Supercomputing and Networking Center (PCSS).}

%
%





\end{document}